\documentclass[%
superscriptaddress, %
twocolumn, %
showpacs,
 amsmath,amssymb,
 aps,
prb,
floatfix, longbibliography ]{revtex4-1}
\usepackage{graphicx}
\usepackage{dcolumn}
\usepackage{bm}
\usepackage{soul,color}

\usepackage[colorlinks,linkcolor=blue,anchorcolor=blue,citecolor=blue,urlcolor=blue]{hyperref}
\usepackage{tabularx}
\usepackage{mathtools}

\newcommand{\mb}[1]{\mathbf{#1}}

\begin{document}

\title{Misalignment instability in magic-angle twisted bilayer graphene on hexagonal boron nitride}

\author{Xianqing Lin}
\email[E-mail: ]{xqlin@zjut.edu.cn}
\affiliation{College of Science,
             Zhejiang University of Technology,
             Hangzhou 310023, People's Republic of China}
             
\author{Kelu Su}
\affiliation{College of Science,
             Zhejiang University of Technology,
             Hangzhou 310023, People's Republic of China}

\author{Jun Ni}
\affiliation{State Key Laboratory of Low-Dimensional Quantum Physics and Frontier Science Center for Quantum Information,
             Department of Physics, Tsinghua University, Beijing 100084,
             People's Republic of China}

\date{\today}

\begin{abstract}
We study the stability and electronic structure of magic-angle twisted bilayer graphene
on the hexagonal boron nitride (TBG/BN).
Full relaxation has been performed for commensurate supercells of the heterostructures with different twist angles ($\theta'$)
and stackings between TBG and BN. We find that the slightly misaligned configuration with  $\theta' = 0.54^\circ$ and the
AA/AA stacking has the globally lowest total energy due to the constructive interference of the moir\'{e} interlayer potentials and thus the
greatly enhanced relaxation in its $1 \times 1$ commensurate supercell.
Gaps are opened at the Fermi level ($E_F$) for small supercells with the stackings that enable strong breaking of the $C_2$
symmetry in the atomic structure of TBG.
For large supercells with $\theta'$ close to those of the $1 \times 1$ supercells, the broadened flat bands can still be resolved from the
spectral functions. The $\theta' = 0.54^\circ$ is also identified as a critical angle for the evolution of the electronic structure with $\theta'$,
at which the energy range of the mini-bands around $E_F$ begins to become narrower with increasing $\theta'$ and their gaps from the dispersive bands
become wider.
The discovered stablest TBG/BN with a finite $\theta'$ of about $0.54^\circ$ and its gapped flat bands
agree with recent experimental observations.
\end{abstract}

\pacs{%
}



\maketitle


\section{Introduction}

The recently realized magic-angle twisted bilayer graphene (TBG) has inspired
great interest in exploring its peculiar
electronic structure
\cite{Cao2018,cao2018unconventional,EmergentSharpe605,lu2019superconductors,uri2020mapping}.
Superconductivity and correlated-insulator
phases associated with the low-energy flat bands have been observed
in TBG with twist angles around the first magic angle ($\theta_m$) of about 1.1$^\circ$
\cite{Cao2018,cao2018unconventional,EmergentSharpe605,lu2019superconductors,uri2020mapping,
Bistritzer12233,Fang2016,OriginPhysRevLett.122.106405}.
In the TBG devices, the hexagonal boron nitride (BN) not only acts as the ideal atomically flat
van der Waals substrates but also can facilitate the realization of
the quantized anomalous Hall (QAH) effect in TBG nearly aligned with BN (TBG/BN)\cite{Intrinsic2020Serlin}.
Due to the lattice-constant mismatch between graphene and BN, the largest moir\'{e} supercell of graphene on BN occurs when the layers are
perfectly aligned\cite{Effect2015Oct,Origin2015Feb,Moire2017Aug}.
In contrast, the twist angle ($\theta'$) between TBG and BN in the experimental devices with the QAH effect
were observed to be about 0.6$^\circ$\cite{Intrinsic2020Serlin}.
The energetics mechanism behind this remains to be revealed.
For the pristine TBG with $\theta_m$, the flat bands around the Fermi level ($E_F$) already have the minimum widths.
The effects of BN on the electronic structure associated with the flat bands in TBG/BN can be beyond the perturbation regime
as BN induces both the direct modification of the Hamiltonian and strong structural deformation in the graphene layer adjacent
to BN that break the $C_2$ symmetry of TBG\cite{Lin2020Symmetry}.
Previous studies only focused on specific configurations of TBG/BN with the minimum commensurate supercells or only considered
the rigid superlattices\cite{Mechanism2020nick,Twisted209zhang,Spontaneous2019liu,zhang2020correlated,Lin2020Symmetry,Cea2020}.
Therefore, it is important to systematically explore the evolution of the energetic and electronic properties of fully relaxed TBG/BN with $\theta'$
to understand the stability and band structures of the experimentally observed configurations.

\begin{figure*}[t]
\includegraphics[width=1.7\columnwidth]{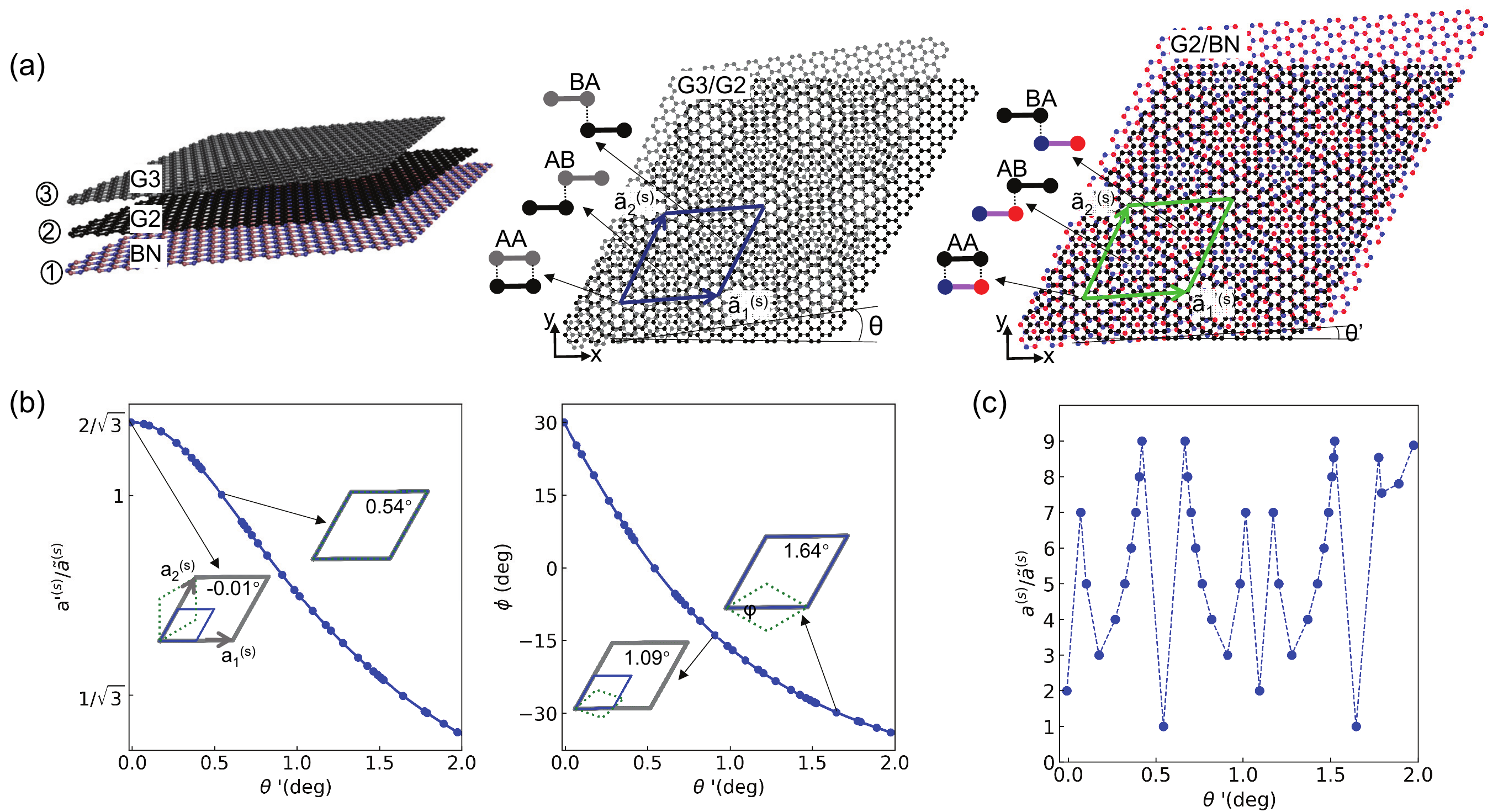}
\caption{(Color online) The geometry of the commensurate supercells of TBG/BN.
(a) The schematic view of the TBG/BN trilayer. The top graphene layer (G3) and the bottom BN layer are rotated by
$\theta$ and $\theta'$ counterclockwise with respect to the fixed middle graphene layer (G2), respectively.
The B and N atoms are represented by blue and red circles, respectively.
The high-symmetry local stackings between G3 and G2 and those between G2 and BN are labeled.
$\mb{\tilde{a}_j^{(s)}}$ ($j=1,2$) are the basis vectors of the moir\'{e} superlattice in G3/G2, and
$\mb{\tilde{a}_j^{'(s)}}$ are those in G2/BN.
(b) Variations of the length ($\tilde{a}^{'(s)}$) of $\mb{\tilde{a}_j^{'(s)}}$ and the angle ($\phi$) from $\mb{\tilde{a}_1^{'(s)}}$ to $\mb{\tilde{a}_1^{(s)}}$
as a function of $\theta'$ for $\theta = 1.08^\circ$.
$\tilde{a}^{(s)}$ is the length of $\mb{\tilde{a}_j^{(s)}}$.
The filled circles represent the strictly commensurate supercells.
The four small supercells are displayed schematically and their $\theta'$ are labeled.
The gray, blue, and green lines represent the supercell of TBG/BN, the moir\'{e} cell of
G3/G2, and that of G2/BN, respectively. $\mb{a_j^{(s)}}$ are the supercell vectors.
(c) The length ($a^{(s)}$) of $\mb{a_j^{(s)}}$ for the supercells at different $\theta'$.
\label{fig1}}
\end{figure*}

Here full relaxation of the commensurate supercells of magic-angle TBG on BN with different $\theta'$ and stacking configurations is performed and
their electronic structures are acquired
based on the effective Hamiltonian taking into account the relaxation effect and the full moir\'{e} Hamiltonian
induced by BN. We find that the slightly misaligned configuration with  $\theta' = 0.54^\circ$ and the
AA/AA stacking has the globally lowest total energy due to the constructive interference of the moir\'{e} interlayer potentials,
and gaps are opened at $E_F$ for small supercells with the stacking that enables strong breaking of the $C_2$
symmetry in the atomic structure of TBG.
The $\theta' = 0.54^\circ$ is also demonstrated to be a critical angle for the evolution of the electronic structure with $\theta'$,
at which the energy range of the mini-bands around $E_F$ begins to become narrower with increasing $\theta'$ and their gaps from the dispersive bands
become wider.

\section{Commensurate supercells of TBG/BN}

We study the trilayer heterostructures with the top magic-angle TBG nearly aligned with
the bottom BN layer. The top graphene layer (G3) and the bottom BN layer are rotated by
$\theta$ and $\theta'$ counterclockwise respectively with respect to the fixed middle graphene layer (G2),
as shown schematically in Fig. 1(a). Due to the relative twist between the adjacent layers and the lattice-constant
mismatch between graphene and BN, double moir\'{e} superlattices are formed in TBG/BN.
As the moir\'{e} superlattices in G3 on G2 (G3/G2) and in G2 on BN (G2/BN) generally have different sizes and orientations,
it appears that the two superlattices may be completely incommensurate.
However, careful examination of the continuous variation of the superlattice in G2/BN with $\theta'$ demonstrates that there exist
a series of
commensurate supercells in TBG/BN which are accessible to the calculations of their relaxed atomic and electronic structures.

The unit cell of the fixed G2 layer is spanned by $\mb{a_1} = a(\sqrt{3}/2, -1/2)^\mathrm{T}$ and
$\mb{a_2} = a(\sqrt{3}/2, 1/2)^\mathrm{T}$, where $a = 2.447$ {\AA} and the superscript T denotes matrix transposition. The cell vectors of the BN layer
become
$\mb{a'_j} = S \mb{a_j}$ ($j$ = 1, 2), where the transformation matrix
\begin{equation}
S = \frac{1}{1 + \epsilon} \left( {\begin{array}{*{20}{c}}
 \cos\theta'  & -\sin\theta'   \\
 \sin\theta' & \cos\theta'
\end{array}} \right)
\end{equation}
and $\epsilon$ is the lattice-constant mismatch between graphene and BN with
the \emph{ab-initio} value of $-1.70\%$.
In the G3 layer, the unit cell is spanned by $T_{\theta} \mb{a_j}$,
where $T_{\theta}$ denotes the counterclockwise rotation by $\theta$.

We consider the strictly periodic moir\'{e} superlattice in G3/G2 with $\theta = 1.0845^\circ$ which is
closest to the experimentally observed $\theta_m$.
This hexagonal superlattice is spanned by the basis vectors
$\mb{\tilde{a}_1^{(s)}} = 30 \mb{a_1} + 31 \mb{a_2}$ and $\mb{\tilde{a}_2^{(s)}} = T_{60^\circ}\mb{\tilde{a}_1^{(s)}} = -31 \mb{a_1} + 61 \mb{a_2}$.
In G2/BN, the spanning vectors of the moir\'{e} superlattices
are taken as $(S^{-1} - I)\mb{\tilde{a}_1^{'(s)}} = -\mb{a_2}$ and $(S^{-1} - I)\mb{\tilde{a}_2^{'(s)}} = \mb{a_1}-\mb{a_2}$ with
$\mb{\tilde{a}_2^{'(s)}} = T_{60^\circ} \mb{\tilde{a}_1^{'(s)}}$.
Starting from the aligned TBG on BN, commensurate supercells of the double superlattices arise
when rotating TBG relative to BN.
For $\theta' = 0^\circ$, the length ($\tilde{a}^{'(s)}$) of $\mb{\tilde{a}_1^{'(s)}}$ is larger than that ($\tilde{a}^{(s)}$) of  $\mb{\tilde{a}_1^{(s)}}$,
while $\tilde{a}^{'(s)}$ decreases with $\theta'$ and becomes equal to $\tilde{a}^{(s)}$ at $\theta' \simeq 0.54^\circ$, as shown in Fig. 1(b).
More importantly, $\mb{\tilde{a}_1^{'(s)}}$ rotates clockwise with increasing $\theta'$
and the angle ($\phi$) from $\mb{\tilde{a}_1^{(s)}}$
to $\mb{\tilde{a}_1^{'(s)}}$ changes from $30^\circ$ at $\theta' = 0^\circ$ to zero at $\theta' \simeq 0.54^\circ$.
Then the two superlattices coincide at $\theta' \simeq 0.54^\circ$ with the minimum commensurate supercell.
At larger $\theta'$, $\tilde{a}^{'(s)}$ becomes smaller than $\tilde{a}^{(s)}$ and the sign of $\phi$ changes.

\begin{figure*}[t]
\includegraphics[width=1.7\columnwidth]{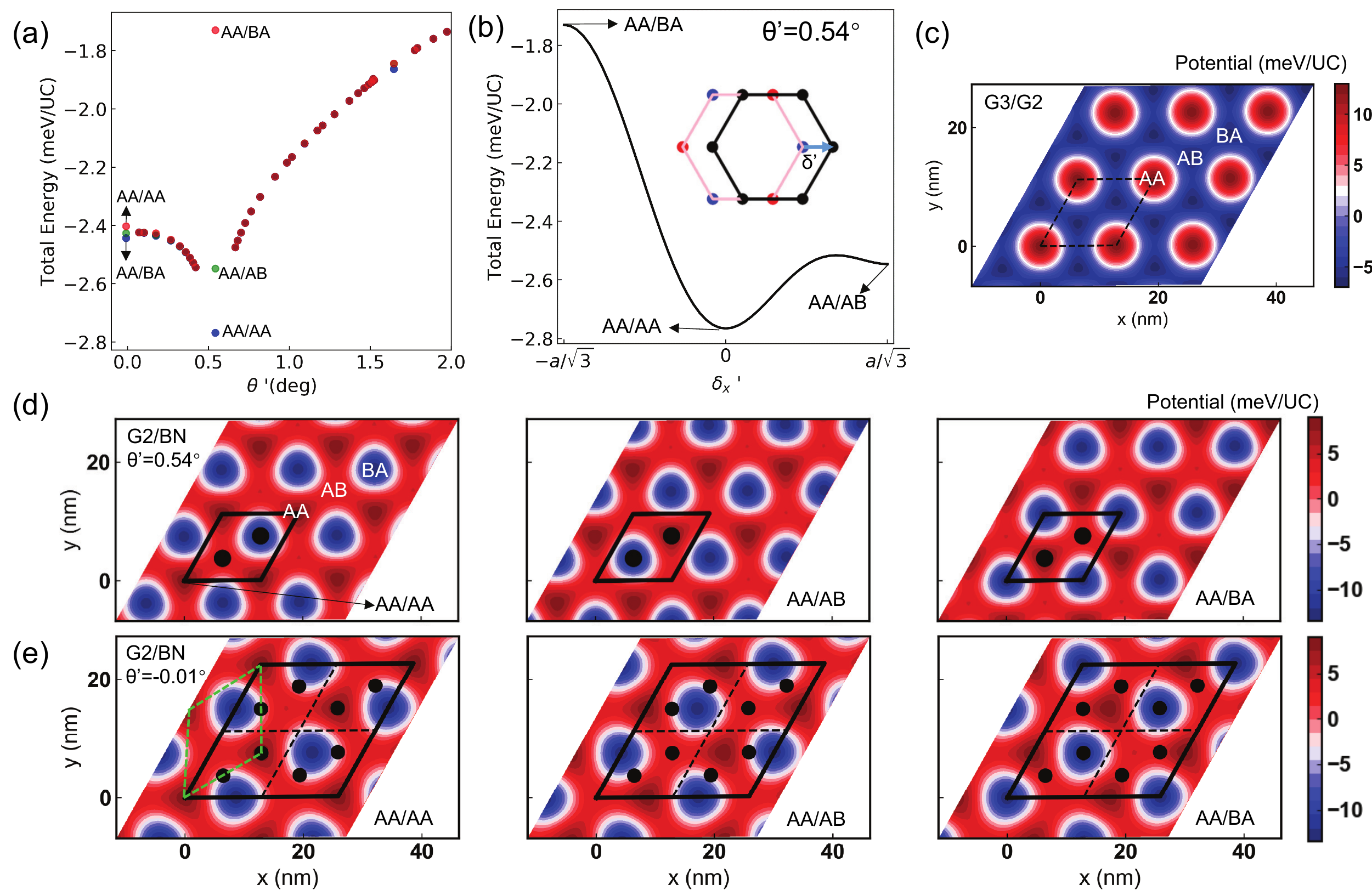}
\caption{(Color online) The energetics of TBG/BN.
(a) The total energies ($E_{tot}$) of the supercells with different high-symmetry stackings and $\theta'$.
The blue, green, and red circles represent the configurations with lowest, middle, and highest $E_{tot}$ at each $\theta'$, respectively.
$E_{tot}$ is in units of meV per area of a graphene unit cell (UC).
(b) $E_{tot}$ of the $1 \times 1$ supercell at $\theta' = 0.54^\circ$ with continuously varying stacking, which is indicated by
the local shift vector $\bm{\delta'}$ between G2 and BN at the origin as seen in the inset.
$\bm{\delta'}$ is taken to be along the $x$ direction.
(c) The spatial distribution of the local interlayer interaction potential of
the rigid superlattice of G3/G2.
(d) and (e) The interlayer potentials of the rigid superlattices of G2/BN for different stackings of the $1 \times 1$ supercell at $\theta' = 0.54^\circ$ (d)
and those for the $2 \times 2$ supercell at $\theta' = -0.01^\circ$ (e). The black solid lines, black dashed lines, and green dashed lines represent
the supercell of TBG/BN, the moir\'{e} cell of
G3/G2, and that of G2/BN, respectively.
The black filled circles denote the positions of the AB and BA stackings between G3 and G2 in one supercell.
\label{fig2}}
\end{figure*}

All the possible hexagonal commensurate supercells are obtained as follows.
The basis vector of the supercell $\mb{a_1^{(s)}} =  n_1 \mb{\tilde{a}_1^{(s)}} + n_2 \mb{\tilde{a}_1^{(s)}} $ satisfies
\begin{equation}
(S^{-1} - I)\mb{a_1^{(s)}} = m_1 (-\mb{a_2}) + m_2 (\mb{a_1}-\mb{a_2})
\end{equation}
with integer values of $n_j$ and $m_j$ so that it is also a lattice vector of the superlattice
in G2/BN. The other basis vector is $\mb{a_2^{(s)}} = T_{60^\circ}\mb{a_1^{(s)}} =  -n_2 \mb{\tilde{a}_1^{(s)}} + (n_1 + n_2) \mb{\tilde{a}_1^{(s)}}$.
The two superlattices coincide when
$(S^{-1} - I)\mb{\tilde{a}_1^{(s)}} = -\mb{a_2}$, which gives $\epsilon = -1.6437\%$ and $\theta' = 0.5423^\circ$.
This $\epsilon$ is just close to the \emph{ab-initio} value of $-1.70\%$.
For other $\theta'$ from 0$^\circ$ to 2$^\circ$, the
$\mb{a_1^{(s)}}$ satisfying Eq. (2) exactly is searched numerically by varying $\epsilon$
around $-1.6437\%$ very slightly. The
obtained strictly periodic supercells are listed in Table SI in the Supplemental Material (SM)\cite{SM}.
We note that to study the energetic stabilities of these supercells, the differences between $\epsilon$
should be extremely small, which are all smaller than $0.026\%$ here.
For most cases in Table SI, the supercell vector can be expressed as $\mb{a_1^{(s)}} = n \mb{\tilde{a}_1^{(s)}}$
with $n = 1\sim9$.
Since these supercells are comprised of $n^2$ TBG moir\'{e} cells,
they are denoted by $n \times n$. Their
structural parameters can be computed analytically as given in the SM.

Figure 1(c) shows the length ($a^{(s)}$) of the supercell basis vectors as a function of $\theta'$.
Four cases have $a^{(s)} \leq 2 \tilde{a}^{(s)}$, including the one with almost perfect alignment between G2 and BN, as shown in Fig. 1(b).
The $a^{(s)}$ of the largest considered supercell reaches $9 \tilde{a}^{(s)}$ (116.3 nm), and such a supercell is comprised of more than one million atoms.
The reciprocal space of the TBG/BN supercell is shown schematically in Fig. S1.

The sublattice-A and sublattice-B atoms in a unit cell of graphene are located at
$(\mb{a_1} + \mb{a_2})/3$ and $(2\mb{a_1} + 2\mb{a_2})/3$, respectively.
In BN, the lattices formed by the boron and nitrogen atoms are labeled as sublattice-A and
sublattice-B, respectively.
Besides the relative twist between BN and TBG, BN can be shifted with respect to TBG forming different stacking configurations.
For the configurations with the $C_3$ symmetry, the stackings of G3/G2 and G2/BN at the origin are one of AA, AB, and BA, as shown in Fig. 1(a), which are used
to denote the stacking between TBG and BN like AA/AA with the left one for G3/G2.
Among the nine possible symmetric stackings, only three are inequivalent.
The local stackings between adjacent layers
vary continuously and are characterized by the relative shift vectors.
At an in-plane position $\mb{r}$ in the rigid superlattice, the shift vector between G3 and G2 is taken to be
$\bm{\delta} = (I - T_{-\theta}) \mb{r} + \bm{\tau}_{32}$, and
that between G2 and BN is given by $\bm{\delta'} = (S^{-1} - I) \mb{r} + \bm{\tau}_{21}$, with
$\bm{\tau}_{32}$ and $\bm{\tau}_{21}$ the shift vectors at the origin.

\section{Energetics of TBG/BN}

\begin{figure}[t]
\includegraphics[width=1.0\columnwidth]{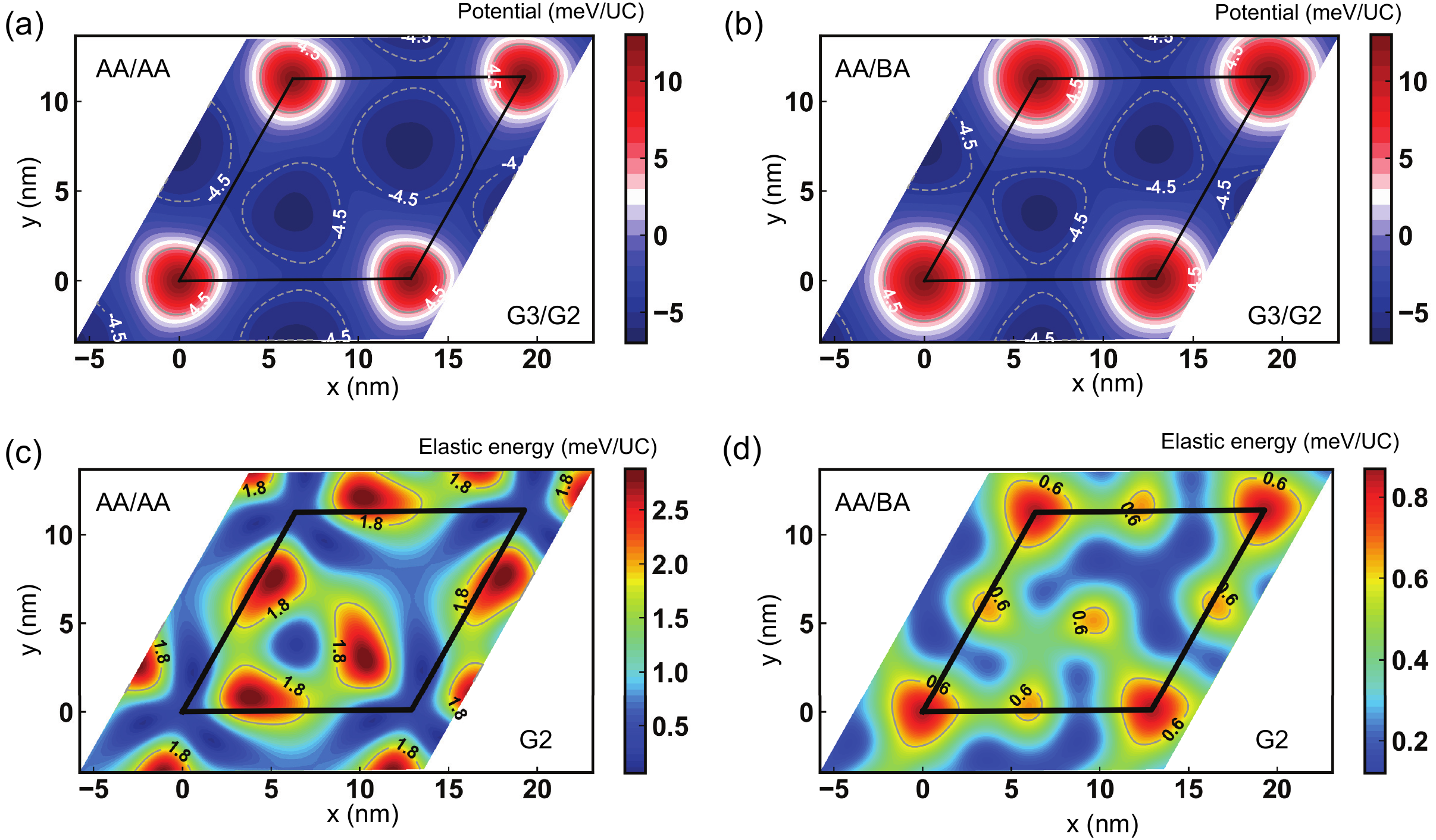}
\caption{(Color online)
The spatial distributions of the interlayer interaction potentials between G3 and G2 (a, b) and the local elastic energy in G2(c, d) for
relaxed TBG/BN with the $1 \times 1$ supercell at $\theta' = 0.54^\circ$ and
the AA/AA and AA/BA stackings.
\label{fig3}}
\end{figure}

\begin{figure*}[t]
\includegraphics[width=1.8\columnwidth]{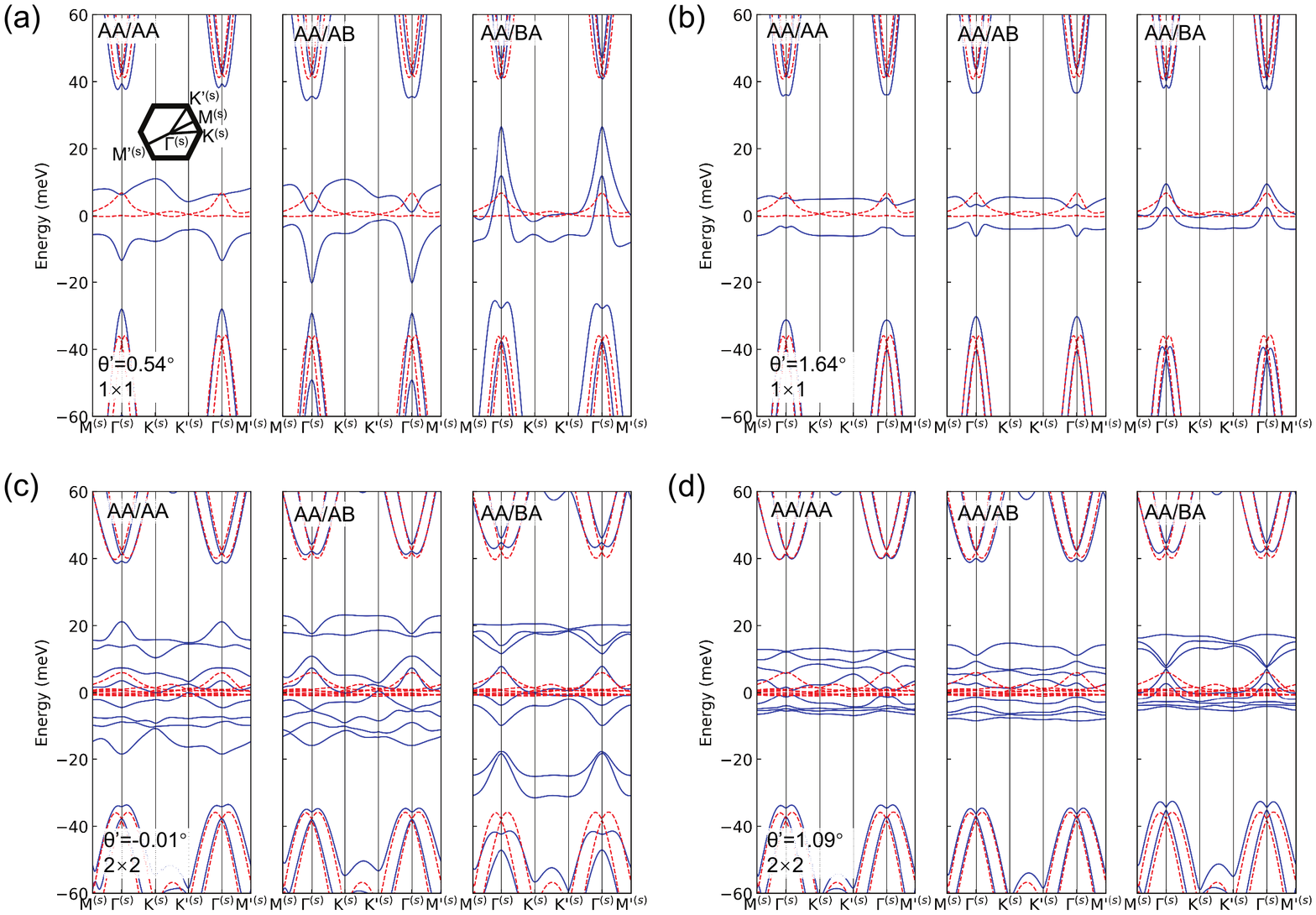}
\caption{(Color online) The band structures of TBG/BN with the $1 \times 1$ supercell at $\theta' = 0.54^\circ$ (a) and $\theta' = 1.64^\circ$ (b) and with the
$2 \times 2$ supercell at $\theta' = -0.01^\circ$ (c) and
$\theta' = 1.09^\circ$ (d) for the three
inequivalent symmetric stackings.
The superimposed bands of the pristine TBG without BN are represented by dashed lines.
The Fermi levels are set to be zero.
\label{fig4}}
\end{figure*}

The rigid moir\'{e} superlattices undergo spontaneous in-plane relaxation due to the energy gain
from the larger domains of energetically favorable local stackings\cite{McEuenBLG13,Uchida2014,Wijk2015,Dai2016,Jain2017,Nam2017,
Gargiulo2018,carr2018relaxation,
Atomicyoo2019atomic,CrucialPhysRevB.99.195419,ContinuumPhysRevB.99.205134}.
Each layer is also corrugated to reach the optimal interlayer distances
of the varying local stackings across the superlattices.
We have performed full
relaxation of TBG/BN with different $\theta'$ and stackings employing the continuum elastic
theory, as detailed in the SM.

The variation of the total energy ($E_{tot}$) of TBG/BN with $\theta'$ and stackings shows
that the energetically stablest configuration occurs at $\theta' = 0.54^\circ$ and has the AA/AA stacking, as shown in Fig. 2(a).
The elastic and interlayer interaction energies that sum to $E_{tot}$ can be seen in Fig. S2.
At $\theta' = 0.54^\circ$ with the $1 \times 1$ supercell, the three stackings have distinct
$E_{tot}$, while the $E_{tot}$ for different stackings at other $\theta'$ are rather similar.
The general trend of $E_{tot}$ with $\theta'$ suggests that $E_{tot}$ reaches minimum
at about $0.54^\circ$ even without strict commensurability.
To realise the stablest AA/AA stacking at $\theta' = 0.54^\circ$, there should not be high energy barriers between the three symmetric
stackings. An energy barrier indeed exists when continuously shifting TBG relative to BN from AA/AB to AA/AA, while it
is just 0.03 meV/UC, much smaller than the difference of $E_{tot}$ between the two symmetric stackings, as shown in Fig. 2(b).
Therefore, TBG/BN tends to have the configuration with $\theta' = 0.54^\circ$ and the AA/AA stacking
when the $\theta'$ of the initial assembled structure is close to this angle and is allowed to be relaxed by, for example,
annealing. Other structures with $\theta'$ rather away from this angle may still exist.
In experiments, TBG/BN with $\theta' \approx 0.6^\circ$ was realized.
Such a slightly larger $\theta'$ than $0.54^\circ$ can be due to a larger $\theta \approx  1.15^\circ$ within the
experimental samples and the ambiguity of the experimental determination of $\theta'$.

The mechanism behind the minimum $E_{tot}$  with $\theta' = 0.54$ and the AA/AA stacking is that
the moir\'{e} interlayer potentials in G3/G2 and G2/BN interfere constructively so that
the most favorable BA and AB stackings in G3/G2 are located at the same position as the most favorable BA stacking
and less favorable AB stacking in G2/BN, as seen in Fig. 2(d).
In contrast, the completely destructive interference of the moir\'{e} potentials for the  AA/AB stacking  at $\theta' = 0.54^\circ$
leads to the highest $E_{tot}$ among all $\theta'$ and stackings.
The interlayer potentials for the almost perfect alignment of TBG on BN ($\theta' = -0.01^\circ$) are also exhibited  in Fig. 2(e) for
comparison. The supercell of these configurations is comprised of four moir\'{e} cells of G3/G2 and three
moir\'{e} cells of G2/BN. Among the three BA-stacked positions of G2/BN in one supercell, two coincide with the BA and AB stackings in G3/G2
for the AA/BA case so that it has a lower energy, while the BA-stacked positions in G2/BN are still located in the vicinity of some
BA and AB stackings in G3/G2 for the other two cases.
Then the $E_{tot}$ of the three stackings are similar at $\theta' = -0.01^\circ$.

The constructive interference of the moir\'{e} potentials
results in the strongest atomic relaxation
for the AA/AA stacking at $\theta' = 0.54$,
in contrast to the suppressed relaxation for the AA/BA stacking, as demonstrated in Fig. 3.
Upon relaxation, the regions with favorable BA-like and AB-like stackings in G3/G2 increase in size, and they are much larger for AA/AA than that
for AA/BA.
The stronger relaxation for AA/AA is also reflected evidently in the spatial distributions of the elastic-energy density, whose highest value for
AA/AA is much higher than that for AA/BA.
When $\theta'$ is away from $0.54^\circ$ with a larger value, the relaxation also becomes much weaker than that at smaller
$\theta'$, as reflected in the more suppressed structural deformation in G2 at larger $\theta'$ (see Fig. S4).

\section{Electronic structure of TBG/BN}

For the relaxed TBG/BN, we have built an effective Hamiltonian $\hat{H}$ for the
moir\'{e} superlattice in G3/G2 by extending the Hamiltonian of $p_z$ orbitals for graphene bilayers
and the effective Hamiltonian of monolayer graphene on BN\cite{Lin2020Symmetry}.
This effective Hamiltonian reads
\begin{eqnarray}
\hat{H} &=& \sum_{n=2}^{3} \sum_{i} \varepsilon_{n, i} c^{\dagger}_{n,i} c_{n,i} +
\sum_{n=2}^{3} \sum_{\left<i,j\right>} t^{(n,n)}_{\left<i,j\right>} (c^{\dagger}_{n,i} c_{n,j} + h.c.)\nonumber \\
  &+& \sum_{i,j} t^{(2,3)}_{i,j} (c^{\dagger}_{2,i} c_{3,j} + h.c.),
\end{eqnarray}
where $c^{\dagger}_{n,i}$ $(n=2,3)$ is the creation and $c_{n,i}$ is the annihilation operator of
a $p_z$-like orbital at the site $i$ in the G$n$ layer and
$\left<i,j\right>$ denotes the intralayer nearest neighbors.
The on-site energies, intralayer and interlayer hopping terms
are represented by $\varepsilon_{n, i}$, $t^{(n,n)}_{i,j}$, and $t^{(2,3)}_{i,j}$, respectively.
These Hamiltonian terms are obtained taking into account the relaxation effect and the full moir\'{e} Hamiltonian
induced by BN, and $\hat{H}$ is diagonalized using
the plane-wave-like basis functions, as detailed in the SM.

\begin{figure*}[tb]
\includegraphics[width=1.7\columnwidth]{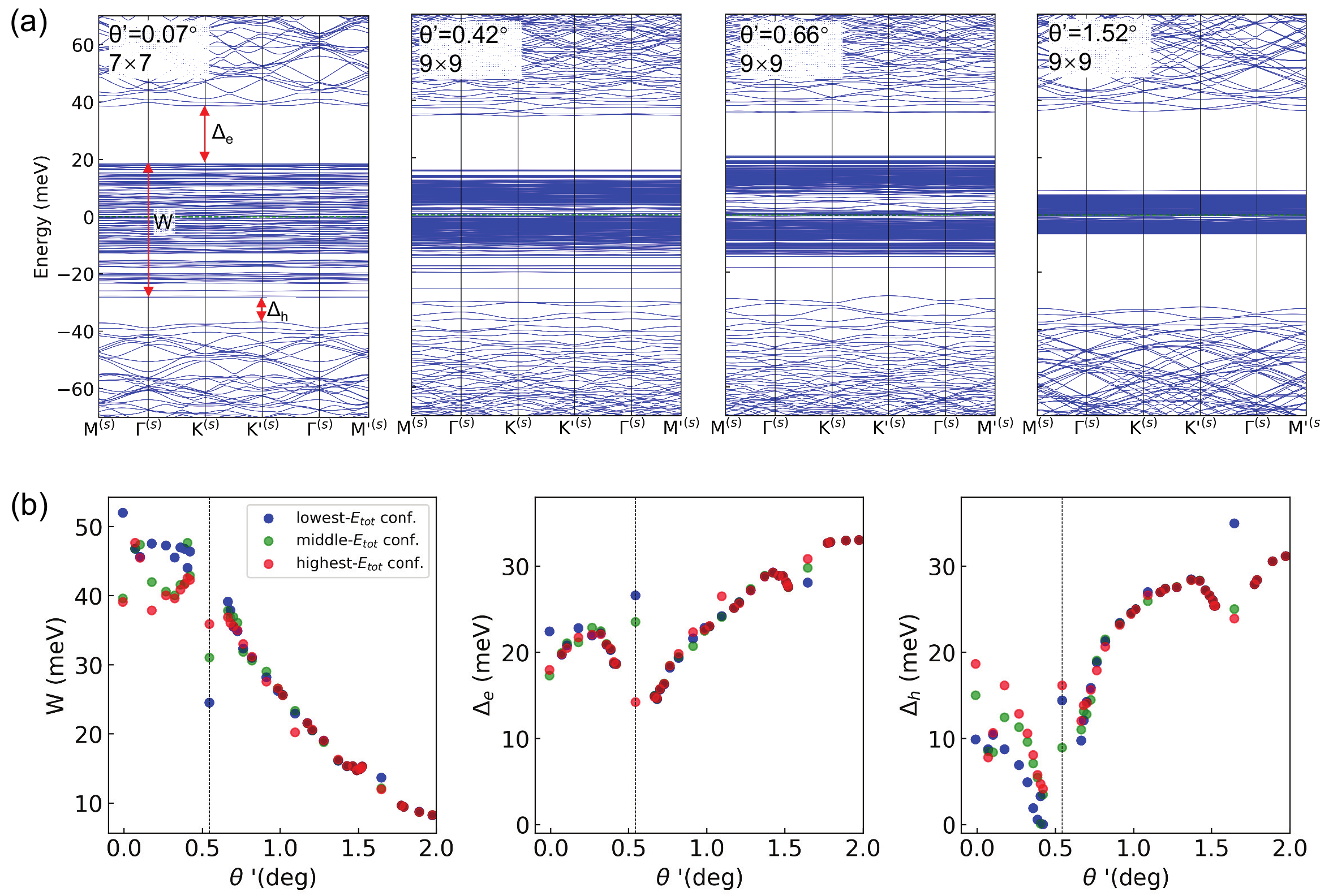}
\caption{(Color online) (a) The band structures of TBG/BN with large supercells at four different $\theta'$.
(b) The energy range ($W$) of the middle mini-bands, the gap ($\Delta_e$) between the dispersive conduction bands and the middle mini-bands, and
the gap ($\Delta_h$) separating the middle mini-bands from the dispersive valence bands for
all considered configurations with different stackings and $\theta'$.
\label{fig5}}
\end{figure*}

The most desirable consequence of the near alignment of TBG with BN for the electronic structure is that
the flat conduction and valence bands can become separated by a gap at $E_F$ due to the broken $C_2$ symmetry in TBG by BN.
Our calculations show that such a gap is only present in systems with small commensurate supercells, as shown in Fig. 4.
Moreover, among the three stackings, the flat bands could be gapped at $E_F$ for the AA/AA and AA/AB stackings, while
they overlap at $E_F$ for AA/BA. For AA/AA and AA/AB, the $C_2$ symmetry is strongly broken
in the relaxed atomic structure due to the evident absence of the inversion symmetry with respect to the origin in the interlayer potential between G2 and BN
[see Fig. 2(d)]. In contrast, the interlayer potential between G2 and BN is roughly inversion symmetric with respect to the origin for AA/AB
so that the $C_2$ symmetry is only weakly broken by the structural relaxation, and the avoided crossings of the flat bands can be attributed to
the electronic contribution of BN.
Although the gap at $E_F$ only arises in specific configurations, it can be readily observed in experiment as the globally stablest system with
$\theta' = 0.54$ and the AA/AA stacking
just has a rather large gap, as shown in Fig. 4(a).

For the $2 \times 2 $ supercells at $\theta' = -0.01^\circ$ and $1.09^\circ$, the mini-bands
on the electron or hole side could also be separated from each other, as shown in Figs. 4(c, d). However, the
band separation depends on the stacking. For example, at $\theta' = -0.01^\circ$, the first two conduction bands are gapped from the higher bands for AA/AA and AA/AB, while
only the lowest conduction band is gapped from the other three overlapped conduction bands for AA/BA.
Such sensitivity of band structures to stackings of small supercells suggests complicated mini-bands in large supercells.

\begin{figure}[tb]
\includegraphics[width=1.0\columnwidth]{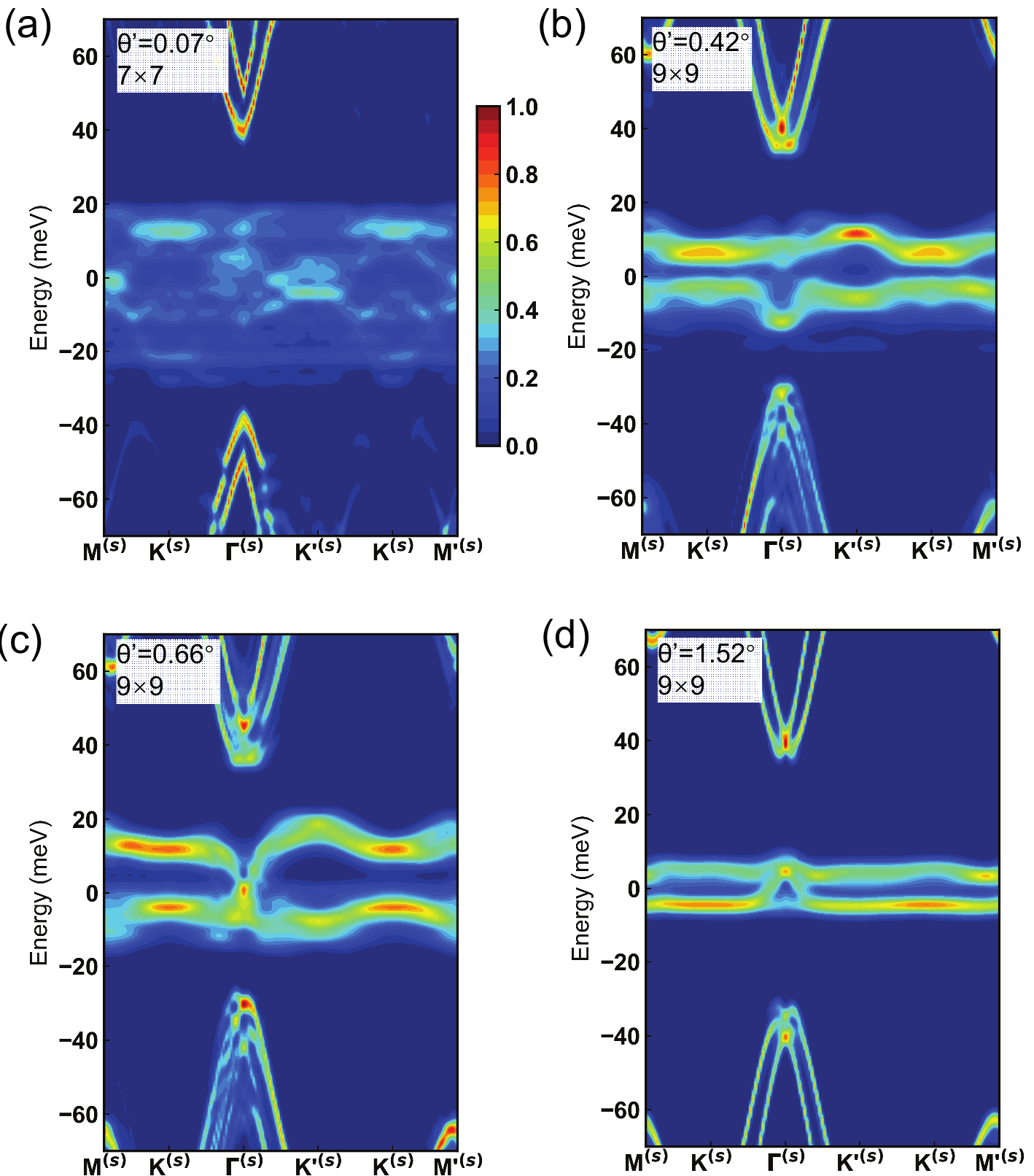}
\caption{(Color online)
The spectral functions unfolded to the BZ of the pristine TBG corresponding to the band structures in Fig. 5(a) for large supercells at four $\theta'$.
The $0.07^\circ$ of $\theta'$ (a) is close to $-0.01^\circ$ with a $2 \times 2$ supercell,
$0.42^\circ$ (b) and $0.66^\circ$ (c) are close to $0.54^\circ$ with a $1 \times 1$ supercell,
and $1.52^\circ$ (d) is close to $1.64^\circ$ also with a $1 \times 1$ supercell.
\label{fig6}}
\end{figure}

Figure 5 shows the band structures of TBG/BN with the  $7 \times 7$ or $9 \times 9$ supercell at four different $\theta'$.
The bands of such large supercells are no longer sensitive to stackings.
Most middle mini-bands around $E_F$ become completely flat lines and are well separated from the dispersive bands, while
no gaps can be observed at $E_F$.
For the two smaller $\theta'$, the energy range ($W$) of the middle mini-bands is rather large, while it
decreases with the two larger $\theta'$. The gap ($\Delta_e$) between the dispersive conduction bands and the middle mini-bands
are all rather large, while the middle mini-bands approach the dispersive valence bands with a vanishing gap ($\Delta_h$) between them at $\theta' = 0.42^\circ$.
The variations of $W$, $\Delta_e$, and $\Delta_h$ with $\theta'$ are exhibited in Fig. 5(b), which clearly infer that
$\theta' = 0.54^\circ$ is the critical angle for the evolution of the electronic structure of TBG/BN with $\theta'$.
At $\theta' < 0.54^\circ$, $W$ remains large and $\Delta_h$ can reach near zero, which can be attributed to
the significant contribution of BN to the structural relaxation and electronic perturbation in TBG as the
moir\'{e} cell of G2/BN is larger than that of G3/G2.
At $\theta' > 0.54^\circ$, $W$ decreases rapidly with $\theta'$ and the increasing $\Delta_h$ becomes rather large.

In order to resolve the effective energy dispersions for large supercells with complicated mini-bands around $E_F$, the
spectral functions unfolded to the BZ of the pristine TBG have been computed, as shown in Fig. 6.
We find that the flat bands corresponding to those of the pristine TBG could still be identified for systems close to the $1 \times 1$
commensurate configurations [see Figs. 6(b-d)], while no energy dispersions could be resolved from the spectral function for the system close to the
$2 \times 2$ configuration [see Fig. 6(a)].
The broadened flat bands could be well separated [see Fig. 6(b)], while they could also overlap in the vicinity of $\Gamma^{(s)}$ [see Figs. 6(c-d)].
These features of the spectral functions may be observed in future ARPES measurements of TBG/BN.

\section{Summary and Conclusions}

A series of commensurate supercells of TBG/BN with magic angle within TBG and with varying $\theta'$ and stackings
between TBG and BN have been constructed.
Full relaxation of the supercells has been performed, which shows that
the energetically stablest configuration is slightly misaligned between TBG and BN with $\theta' = 0.54^\circ$ and has the AA/AA stacking.
This is due to the completely constructive interference of the moir\'{e} interlayer potentials and thus the
greatly enhanced relaxation in the $1 \times 1$ supercell of this configuration.
In contrast, in-plane relaxation can be partially suppressed in the graphene layer adjacent to BN for other configurations.
The band structures of the supercells are acquired
based on the effective Hamiltonian taking into account the relaxation effect and the full moir\'{e} Hamiltonian
induced by BN. As the supercells have a huge number of atoms, the Hamiltonian is diagonalized using the plane-wave-like basis.
Gaps are opened at $E_F$ for small supercells with stackings that enable strong breaking of the $C_2$
symmetry in the atomic structure of TBG.
For large supercells with $\theta'$ close to those of the $1 \times 1$ supercells, the broadened flat bands can still be resolved from the
spectral functions. The $\theta' = 0.54^\circ$ is also identified as a critical angle for the evolution of the electronic structure with $\theta'$,
at which the energy range of the mini-bands around $E_F$ begins to become narrower with increasing $\theta'$ and their gaps from the dispersive bands
become wider.
The discovered stablest TBG/BN with a finite $\theta'$ of about $0.54^\circ$ and its gapped flat bands
agree with recent experimental observations.

\label{Acknowledgments}
\begin{acknowledgments}
We gratefully acknowledge valuable discussions with D. Tom\'anek and
H. Xiong.
This research was supported by
the National Natural Science Foundation of China (Grants No. 11974312 and No. 11774195),
and the National Key Research and Development Program of China(Grant No. 2016YFB0700102).
The calculations were performed on TianHe-1(A) at National Supercomputer Center in Tianjin.
\end{acknowledgments}


%

\end{document}